\documentstyle[12pt]{article}

\def\proof{\medbreak\noindent{\bf Proof}}

\def\theorem #1. #2\par{\medbreak
  \noindent{\tt {\bf Theorem #1.}\enspace}{\sl#2\par}%
  \ifdim\lastskip<\medskipamount \removelastskip\penalty55\medskip\fi}

\def\ww{\wedge\ldots\wedge}
\def\la{\Lambda}
\def\tr{{\rm Tr}}
\def\I{{\cal I}}
\def\R{{\cal R}}
\def\M{{\cal M}}
\def\C{{\cal C}}
\def\L{{\cal L}}
\def\Spin{{\rm Spin}}
\def\det{{\rm det}}
\def\SO{{\rm SO}}
\def\exp{{\rm exp}}
\def\cos{{\rm cos}}
\def\sin{{\rm sin}}

\def\be{\begin{equation}}
\def\ee{\end{equation}}

\begin{document}

\author{N.G.Marchuk}

\title{The Dirac equation vs. the Dirac type
tensor equation}

\maketitle

PACS 04.20.Cv, 11.15

\begin{abstract}We discuss a connection between the Dirac equation for an
electron and the Dirac type tensor equation  with ${\rm U}(1)$ gauge symmetry.
\end{abstract}

\bigskip

In the previous paper \cite{Marchuk:Cimento1}, using results of P.~Dirac \cite{Dirac},
D.~Ivanenko and L.~Landau \cite{Ivanenko}, E.~K\"ahler \cite{Kahler},
F.~G\"ursey \cite{Gursey}, D.~Hestenes \cite{Hestenes}, \cite{Hestenes1},
we present the,
so-called, Dirac type tensor equation. In this paper, developing results of \cite{Marchuk:Cimento1},
we concentrate our attention on a  connection between the
Dirac equation for an electron and the Dirac type tensor equation with
${\rm U}(1)$ gauge symmetry (see formulas (\ref{Dtte},\ref{corr1},\ref{Dirac})).

\bigskip

Let $\R^{1,3}$ be the Minkowski space with coordinates $x^\mu$, with
the metric tensor $\|g_{\mu\nu}\|={\rm diag}(1,-1,-1,-1)$,
$g={\rm det}\|g_{\mu\nu}\|=-1$,
with basis
coordinate vectors $e_\mu$, and with basis covectors
$e^\mu=g^{\mu\nu}e_\nu$. Greek indices run over $0,1,2,3$, Latin indices
run over $1,2,3,4$, and the summation convention over repeating indices
is assumed. Consider a covariant antisymmetric tensor field of rank $k$ in
$\R^{1,3}$
$$
u_{\mu_1\ldots\mu_k}=u_{[\mu_1\ldots\mu_k]},
$$
where square brackets denote the operation of alternation (with division
by $k!$) and $u_{\mu_1\ldots\mu_k}=u_{\mu_1\ldots\mu_k}(x)$ are smooth
real valued functions $\R^{1,3}\to\R$. It is suitable to write this
tensor field as the exterior form
\begin{equation}
\frac{1}{k!}u_{\mu_1\ldots\mu_k}e^{\mu_1}\ww e^{\mu_k}.
\label{ex}
\end{equation}
Under a linear nondegenerate change of coordinates
$$
\acute{x}^\mu=p^\mu_\nu x^\nu,
$$
where $p^\mu_\nu$ are real constants, the components of tensor field
transform as
$$
\acute{u}_{\nu_1\ldots\nu_k}=
q^{\mu_1}_{\nu_1}\ldots q^{\mu_k}_{\nu_k}u_{\mu_1\ldots\mu_k}
$$
and aggregates $e^{\mu_1}\ww e^{\mu_k}$ transform as components of
contravariant tensor field of rank $k$
$$
\acute{e}^{\nu_1}\ww \acute{e}^{\nu_k}=
p^{\nu_1}_{\mu_1}\ldots p^{\nu_k}_{\mu_k} e^{\mu_1}\ww e^{\mu_k},
$$
where
$$
p^\nu_\mu q^\lambda_\nu=\delta^\lambda_\mu,\quad
p^\nu_\mu q_\lambda^\mu=\delta_\lambda^\nu
$$
and $\delta^\mu_\mu=1$, $\delta^\mu_\nu=0$ for $\mu\neq\nu$. That means
the exterior form (\ref{ex}) is an invariant
$$
\frac{1}{k!}u_{\mu_1\ldots\mu_k}e^{\mu_1}\ww e^{\mu_k}=
\frac{1}{k!}\acute{u}_{\nu_1\ldots\nu_k}\acute{e}^{\nu_1}\ww\acute{e}^{\nu_k}.
$$
In this paper we admit  changes of coordinates from the proper
orthohroneous Lorentz group ${\rm SO}^+(1,3)$ only, i.e.,
$$
P^T g P=g,\quad {\rm det}\,P=1,\quad p^0_0>0,
$$
where $P=\|p^\mu_\nu\|$.

Also we consider nonhomogeneous exterior forms
$$
U=\sum^4_{k=0}\frac{1}{k!}u_{\mu_1\ldots\mu_k}e^{\mu_1}\ww e^{\mu_k}.
$$
Denote by $\Lambda$ the set of all such exterior forms and by $\Lambda_k$ the
sets of exterior forms of rank $k$.
The set $\la_0$ is identified with the set of smooth scalar functions $\R^{1,3}\to\R$.
We have
\begin{eqnarray*}
&&\Lambda=\Lambda_0\oplus\cdots\oplus\Lambda_4=\Lambda_{\rm ev}\oplus\Lambda_{\rm od},\\
&&\Lambda_{\rm ev}=\Lambda_0\oplus\Lambda_2\oplus\Lambda_4,\quad
\Lambda_{\rm od}=\Lambda_1\oplus\Lambda_3.
\end{eqnarray*}
Elements of $\la_{\rm ev}$ and $\la_{\rm od}$ are called {\it even} and {\it odd}
exterior forms respectively.
At any point $x\in\R^{1,3}$ we may consider the sets
$\Lambda,\Lambda_{\rm ev},\Lambda_{\rm od},
\Lambda_0,\Lambda_1,\Lambda_2,\Lambda_3,\Lambda_4$ as linear spaces of
dimensions 16,8,8,1,4,6,4,1 respectively. Basis elements of $\la$ are
$$
1,\,e^\mu,\,e^{\mu_1}\ww e^{\mu_k},\quad \mu_1<\cdots<\mu_k,
\quad k=2,3,4.
$$
Also we deal with sets $\Lambda^\C,\Lambda_{\rm ev}^\C,\Lambda_{\rm od}^\C,
\Lambda_0^\C,\Lambda_1^\C,\Lambda_2^\C,\Lambda_3^\C,\Lambda_4^\C$ of complex valued
exterior forms.

The exterior product $U,V\to U\wedge V$ of exterior forms is defined in the usual way and
$$
U\wedge V=(-1)^{rs}V\wedge U\in\Lambda_{r+s},\quad\hbox{for}\quad U\in\Lambda_r,\,V\in\Lambda_s.
$$

Consider {\it a Hodge star operator} $\star\,:\,\Lambda_k\to\Lambda_{4-k}$.
By definition, put
$$
\star U=\frac{1}{k!(4-k)!}\epsilon_{\mu_1\ldots\mu_4}u^{\mu_1\ldots\mu_k}
e^{\mu_{k+1}}\ww e^{\mu_4},
$$
where $U\in\la_k$ is from (\ref{ex}), $\epsilon_{\mu_1\ldots\mu_4}$ is the totally antisymmetric
tensor ($\epsilon_{0123}=1$) and
$u^{\mu_1\ldots\mu_k}=g^{\mu_1\nu_1}\ldots g^{\mu_k\nu_k}u_{\nu_1\ldots\nu_k}$.
It can be checked that
$$
\star\star U=(-1)^{k+1} U.
$$

Let us define {\it a central product}\footnote{In a special
case the central product was
invented by H.~Grassmann \cite{Grassmann} in 1877 as an attempt to
unify the exterior calculus (the Grassmann algebra) with the quaternion
calculus. A discussion on that matter see in \cite{Doran}. In some
papers the central product is called a Clifford product.}
 of exterior forms $U,V\to UV$ by the following rules:
\begin{description}
\item[1.] For $U,V,W\in\la,\,\alpha\in\la_0$
\begin{eqnarray*}
&&1U=U1=U,\\
&&(\alpha U)V=U(\alpha V)=\alpha(UV),\\
&&U(VW)=(UV)W,\\
&&(U+V)W=UW+VW;
\end{eqnarray*}
\item[2.] $e^\mu e^\nu=e^\mu\wedge e^\nu+g^{\mu\nu}$;
\item[3.] $e^{\mu_1}\ldots e^{\mu_k}=e^{\mu_1}\ww e^{\mu_k}\quad$ for
$\quad \mu_1<\cdots<\mu_k$.
\end{description}

Note that from the second rule we get the equalities
$e^\mu e^\nu+e^\nu e^\mu=2 g^{\mu\nu}$, which appear in the Clifford algebra.

\theorem 1. ({\rm \cite{Marchuk:Cimento}, p.1278}). The central product
of exterior forms is an exterior form.\par

In other words, the operation of central product maps $\la\times\la$
to $\la$.
\medskip

It can be checked that
\begin{equation}
UV=U\wedge V-\star(U\wedge\star V)
\label{UV}
\end{equation}
for $U\in\la_1,\,V\in\la$. If we formally substitute $U=e^\mu\partial_\mu$
into (\ref{UV}), then we get
$$
e^\mu\partial_\mu V=dV-\star d\star V=(d-\delta)V,
$$
where $d\,:\,\la_k\to\la_{k+1}$ is called {\it a differential} of
exterior form ($d^2=0$), and $\delta=\star d\star\,:\,\la_k\to\la_{k-1}$
is called {\it a codifferential} of exterior form ($\delta^2=0$).

\medskip

By $\ell$ denote {\it a volume form}
$$
\ell=\sqrt{-g}\,e^0\ww e^3=e^0\ww e^3.
$$
The volume form commutes
with all even exterior forms and anticommutes with all odd exterior forms
with respect to (w.r.t.) the central product.

Suppose that exterior forms $H\in\la_1$; $I,K\in\la_2$ are such
that
\begin{eqnarray}
&&H^2=1,\quad I^2=K^2=-1,\quad [H,I]=[H,K]=\{I,K\}=0,\nonumber\\
&&\partial_\mu H=\partial_\mu I=\partial_\mu K=0,\label{HIK:cond}
\end{eqnarray}
where $[H,I]=HI-IH$, $\{I,K\}=IK+KI$. Then the exterior forms
$\ell,H,I,K$ are said to be {\it invariant generators} of $\la$.
In particular, in fixed coordinates the exterior forms
\begin{equation}
\acute{H}=e^0,\quad \acute{I}=-e^1\wedge e^2,\quad
\acute{K}=-e^1\wedge e^3
\label{HIK:acute}
\end{equation}
satisfy (\ref{HIK:cond}) that means
$\ell,\acute{H},\acute{I},\acute{K}$ are invariant generators
of $\la$.

The 16 exterior forms
\begin{eqnarray*}
&&1\in\la_0;\quad
H,\ell HI,\ell HK,\ell HIK\in\la_1;\\
&&I,K,IK,\ell I, \ell K,\ell IK\in\la_2;\\
&&HI,HK,HIK,\ell H\in\la_3;\quad \ell\in\la_4
\end{eqnarray*}
are linear independent at any point $x\in\R^{1,3}$ and they can be
used as basis exterior forms of $\la$.

Let us take the exterior form
$$
t=\frac{1}{4}(1+H)(1-iI)\in\la^\C
$$
such that
$$
t^2=t,\quad Ht=t,\quad It=it.
$$
The equality $t^2=t$ means that $t$ is an idempotent and we may
consider the left ideal
$$
\I(t)=\{Ut\,:\,U\in\la^\C\}\subset\la^\C.
$$
It can be checked that complex dimension of this left ideal is equal
to four. The exterior forms $t_k=F_k t,\quad k=1,2,3,4$, where
\begin{equation}
F_1=1,\quad F_2=K,\quad F_3=-I\ell,\quad F_4=-KI\ell,
\label{Fk}
\end{equation}
are linear independent and can be considered as basis elements
of $\I(t)$.

\medskip

Let us define operations of conjugation $*$
and Hermitian conjugation $\dagger$,
which map $\la_k\to\la_k$ or $\la_k^\C\to\la_k^\C$. For $U\in\la_k^\C$
\begin{eqnarray*}
U^* &:=& (-1)^{\frac{k(k-1)}{2}}\bar{U},\\
U^\dagger &:=& HU^*H,
\end{eqnarray*}
where  $\bar{U}$ is the exterior form with complex
conjugated components (if $U\in\la$, then $\bar{U}=U$).
We see that
$$
(UV)^*=V^*U^*,\quad U^{**}=U,\quad (UV)^\dagger=V^\dagger U^\dagger,
\quad U^{\dagger\dagger}=U
$$
for $U,V\in\la$ or $\la^\C$.

\medskip

Let us define {\it a trace} of exterior form as the linear operation
$\tr\,:\,\la\to\la_0$ such that
$\tr(1)=1$ and $\tr(e^{\mu_1}\ww e^{\mu_k})=0$ for $k=1,2,3,4$.
It is easy to prove that
$$
\tr(UV-VU)=0\quad\hbox{for}\quad U,V\in\la.
$$

\medskip

Now we may define an operation
$(\,\cdot\,,\,\cdot\,)\,:\,\I(t)\times\I(t)\to\la_0^\C$ by the
formula
$$
(U,V)=4\,\tr(UV^\dagger)\quad\hbox{for}\quad U,V\in\I(t).
$$
This operation has all properties of Hermitian scalar product
\begin{eqnarray*}
&&(\alpha U,V)=\alpha(U,V),\quad (U,V)=(\overline{V,U}),\quad
(U+W,V)=(U,V)+(W,V),\\
&&(U,U)>0\quad\hbox{for}\quad U\neq0,
\end{eqnarray*}
where $U,V,W\in\I(t)$, $\alpha\in\la_0^\C$.
This scalar product converts the left ideal $\I(t)$ into the four
dimensional unitary space with orthonormal basis
$t_k=t^k,\,(k=1,2,3,4)$
$$
(t_k,t^n)=\delta_k^n.
$$

\theorem 2 ({\rm  \cite{Marchuk:Cimento1}, Theorem 4}).
If $\Phi\in\I(t)$ is given and
$\Psi\in\la_{\rm ev}$ is unknown even exterior form, then the equation
$$
\Psi t=\Phi
$$
has a unique solution
$$
\Psi=F_k(\alpha^k+\beta^k I),
$$
where  $\Phi$ has the form
$$
\Phi=(\alpha^k+i\beta^k)t_k
$$
and $F_k$ are defined in
(\ref{Fk}).\par

\medskip

This theorem establishes the one-to-one correspondence between $\la_{\rm ev}$
and $\I(t)$.

\medskip

Let $\M(4,\C)$ be the algebra of $4\!\times\!4$-matrices with complex
valued elements. We define a map $\gamma\,:\,\la\to\M(4,\C)$ with
the aid of equalities
$$
Ut_k=\gamma(U)^n_k t_n,\quad k=1,2,3,4;\quad U\in\la.
$$
Here $\gamma(U)^n_k$ are elements of a four dimensional square matrix
$\gamma(U)$ (an upper index enumerate rows and a lower index enumerate
columns of a matrix). If $\partial_\mu U=0$, then elements of
matrix $\gamma(U)$ are constants. Otherwise they are smooth functions
$\R^{1,3}\to \C$. It is easily shown that
$$
\gamma(UV)=\gamma(U)\gamma(V),\quad
\gamma(U+V)=\gamma(U)+\gamma(V),\quad
\gamma(\alpha U)=\alpha\gamma(U)
$$
for $U,V\in\la$, $\alpha\in\la_0$. Hence the map $\gamma$ is a
matrix representation of $\la$ such that a central product of
exterior forms $UV$ corresponds to the product of matrices
$\gamma(U)\gamma(V)$. This map depends on invariant generators
$\ell,H,I,K$. In particular, if we take invariant generators
(\ref{HIK:acute}), then we get the following well known (Dirac)
representation of matrices $\gamma^\mu=\gamma(e^\mu)$:
\begin{eqnarray*}
\gamma^0&=&\pmatrix{1 &0 &0 & 0\cr
                  0 &1 & 0&0 \cr
                  0 &0 &-1&0 \cr
                  0 &0 &0 &-1},\quad
\gamma^1=\pmatrix{0 &0 &0 &-1\cr
                  0 &0 &-1&0 \cr
                  0 &1 &0 &0 \cr
                  1 &0 &0 &0},\\
\gamma^2&=&\pmatrix{0 &0 &0 & i\cr
                  0 &0 &-i&0 \cr
                  0 &-i&0 &0 \cr
                  i &0 &0 &0},\quad
\gamma^3=\pmatrix{0 &0 &-1& 0\cr
                  0 &0 & 0&1 \cr
                  1 &0 &0 &0 \cr
                  0 &-1&0 &0}.
\end{eqnarray*}

Now we may write down an equation, which we call {\it a Dirac type
tensor equation}
\begin{equation}
(d-\delta)\Psi+A\Psi I+m\Psi HI=0,
\label{Dtte}
\end{equation}
where an even exterior form $\Psi\in\la_{\rm ev}$ is interpreted as a
wave function of an electron, a 1-form $A=a_\mu e^\mu\in\la_1$ is
identified with a potential of electromagnetic field, exterior forms
$H,I$ are defined in (\ref{HIK:cond}), and $m$ is a real nonnegative constant (the electron mass).
Let us take matrices $\gamma^\mu=\gamma(e^\mu)$, defined with the aid of the map $\gamma$,
and scalar functions
$\psi^k\,:\,\R^{1,3}\to\C$ ($\psi^k\in\la_0,\,k=1,2,3,4$) defined by the formula
\begin{equation}
\Psi t=\psi^k t_k.
\label{corr1}
\end{equation}

\theorem 3. Formula (\ref{corr1}) gives one-to-one correspondence
between solutions $\Psi\in\la_{\rm ev}$ of the Dirac type tensor equation
(\ref{Dtte}) and solutions $\psi=(\psi^1\,\psi^2\,\psi^3\,\psi^4)^T$
of the Dirac equation
\begin{equation}
\gamma^\mu(\partial_\mu\psi+i a_\mu\psi)+im\psi=0.
\label{Dirac}
\end{equation}
\par

\proof. Let $\Psi\in\la_{\rm ev}$ be a solution of the Dirac type tensor
equation. Substituting $e^\mu\partial_\mu$ for $d-\delta$ in
(\ref{Dtte}), multiplying both sides of (\ref{Dtte}) from right by
$Ht$, and using the relations $Ht=t$, $It=it$, $\partial_\mu t=0$, we
get
\begin{eqnarray*}
0&=&(e^\mu(\partial_\mu\Psi+a_\mu\Psi I)+m\Psi HI)Ht\\
&=&e^\mu(\partial_\mu(\Psi t)+a_\mu(\Psi t)i)+m(\Psi t)i\\
&=&e^\mu(\partial_\mu(\psi^k t_k)+a_\mu(\psi^k t_k)i)+m(\psi^k t_k)i\\
&=&(e^\mu t_k)(\partial_\mu\psi^k+a_\mu\psi^k i)+m(\psi^k t_k)i\\
&=&\gamma(e^\mu)^n_k t_n(\partial_\mu\psi^k+a_\mu\psi^k i)+m(\psi^n t_n)i\\
&=&(\gamma(e^\mu)^n_k(\partial_\mu\psi^k+a_\mu\psi^k i)+m\psi^n i)t_n.
\end{eqnarray*}
Since $\{t_n\}$ is an orthonormal basis of $\I(t)$, it follows that
$$
\gamma(e^\mu)^n_k(\partial_\mu\psi^k+a_\mu\psi^k i)+m\psi^n i=0,
\quad n=1,2,3,4.
$$
These equalities are equivalent to (\ref{Dirac}). Hence
$\psi=(\psi^1\,\psi^2\,\psi^3\,\psi^4)^T$, where $T$ denote transposition,
is a solution of the Dirac equation.

\medskip

Conversely, suppose that scalar complex valued functions $\psi^k$ are such that
the column $\psi=(\psi^1\,\psi^2\,\psi^3\,\psi^4)^T$ satisfies equation (\ref{Dirac}).
By Theorem 2 there exists a unique solution $\Psi\in\la_{\rm ev}$ of the equation
(\ref{corr1}). Arguing as above but in inverse order, we see that the exterior form
$$
\Omega=(e^\mu(\partial_\mu\Psi+a_\mu\Psi I)+m\Psi HI)H\in\la_{\rm ev}
$$
satisfy equality
$$
\Omega t=0.
$$
By Theorem 2 we get $\Omega=0$. This means that the exterior form $\Psi\in\la_{\rm ev}$
satisfies the Dirac type tensor equation (\ref{Dtte}).
This completes the proof.

\medskip

For the sequel we need a set of even exterior forms
$$
\Spin(1,3)=\{ S\in\la_{\rm ev}\,:\,S^*S=1,\,\partial_\mu S=0\}.
$$
This set can be considered as a group w.r.t. the central product.
It can be shown that if $U\in\la_k$ and $S\in\Spin(1,3)$, then
$S^*US\in\la_k$. In particular,
$$
S^*e^\mu S=p^\mu_\nu e^\nu,
$$
where $p^\mu_\nu$ are real constants and the matrix $P=\|p^\mu_\nu\|$
is such that
\begin{equation}
P^T gP=g,\quad \det P=1,\quad p^0_0>0.
\label{P}
\end{equation}
Now we may consider a change of coordinates
\begin{equation}
x^\mu\to\acute{x}^\mu=p^\mu_\nu x^\nu,\quad
e^\mu\to\acute{e}^\mu=p^\mu_\nu e^\nu,
\label{X}
\end{equation}
which associated with the exterior form $S\in\Spin(1,3)$.
According to the formulas (\ref{P}), this change of coordinates is
from the group $\SO^+(1,3)$.

Conversely, if we take any change of coordinates (\ref{X}) from
the group $\SO^+(1,3)$, i.e., $p^\mu_\nu$ satisfy (\ref{P}), then
there exists a unique pair of exterior forms $\pm S\in\Spin(1,3)$
such that
$$
p^\mu_\nu e^\nu=S^* e^\mu S.
$$

We claim that correspondence (\ref{corr1}) between the Dirac type
tensor equation (\ref{Dtte}) and the Dirac equation (\ref{Dirac})
is the same in any coordinates $\acute{x}^\mu$ such that
transformation $x^\mu\to\acute{x}^\mu$ is from the proper
orthohroneous Lorentz group $\SO^+(1,3)$. Indeed, consider a change
of coordinates (\ref{X}) from the group $\SO^+(1,3)$, which
associated with the exterior form $S\in\Spin(1,3)$. The exterior
forms $\Psi, A,H,I$ from (\ref{Dtte}) are invariants and the
operators $d,\delta$ are invariant under this change of coordinates.
Therefore the Dirac type tensor equation has the same form in
coordinates $\acute{x}^\mu$. Consider relation (\ref{corr1}). As
$\Psi,t,t_k$ are exterior forms, i.e., invariants, the functions
$\psi^k\,:\,\R^{1,3}\to\C$ must be invariants too. Let us write
the Dirac equation (\ref{Dirac}) in coordinates $\acute{x}^\mu$
\begin{equation}
\acute{\gamma}^\mu(\acute{\partial}_\mu\psi+i\acute{a}_\mu\psi)+
i m\psi=0,
\label{Dirac:prime}
\end{equation}
where
\begin{eqnarray*}
&&\acute{\partial}_\mu=\frac{\partial}{\partial\acute{x}^\mu}=
q^\nu_\mu\partial_\nu,\quad
\acute{a}_\mu=q^\nu_\mu a_\nu,\quad
q^\nu_\mu p^\mu_\lambda=\delta^\nu_\lambda,\\
&&\acute{\gamma}^\mu=\gamma(\acute{e}^\mu)=\gamma(p^\mu_\nu e^\nu)
=\gamma(S^*e^\mu S)=\gamma(S^*)\gamma(e^\mu)\gamma(S)=
R^{-1}\gamma^\mu R,\\
&&R=\gamma(S).
\end{eqnarray*}
Substituting $R^{-1}\gamma^\mu R$ for $\acute{\gamma}^\mu$ in
(\ref{Dirac:prime}), we get
\begin{equation}
R^{-1}\gamma^\mu R(\acute{\partial}_\mu\psi+i\acute{a}_\mu\psi)+
i m\psi=0,
\label{D1}
\end{equation}
or, equivalently,
\begin{equation}
\gamma^\mu(\acute{\partial}_\mu(R\psi)+i\acute{a}_\mu(R\psi))+
i m(R\psi)=0.
\label{D2}
\end{equation}
Thus, postulating relation (\ref{corr1}) between the Dirac type
tensor equation and the Dirac equation, we arrive at the
conventional transformation rule for the Dirac equation (\ref{D2})
under changes of coordinates from the group $SO^+(1,3)$.
Let us recall that in the first paper on a theory of electron
\cite{Dirac} P.~A.~M.~Dirac proves covariance of his equation
(\ref{Dirac}) assuming transformation rule (\ref{D1}), i.e., the column
$\psi$ is invariant and $\gamma$-matrices are transform according to the
rule $\gamma^\mu\to R^{-1}\gamma^\mu R$. Later it became conventional
to prove covariance of the Dirac equation assuming transformation rule
(\ref{D2}), i.e., $\gamma$-matrices are invariant and the column $\psi$
transforms according to the rule $\psi\to R\psi$. Columns $\psi$
with such transformation property are called {\it Dirac spinors} or
{\it bispinors}.

Further, let us write the Dirac type tensor equation together with the
Maxwell equations (Quantum Electrodynamics equations)
\begin{eqnarray}
&&H^2=1,\quad I^2=-1,\quad [H,I]=0,\quad \partial_\mu H=\partial_\mu I=0,
\label{main1}\\
&&(d-\delta)\Psi+A\Psi I+m\Psi HI=0,\label{main2}\\
&&dA=F,\label{main3}\\
&&\delta F=\alpha J,\label{main4}\\
&&J=\Psi H\Psi^*,\label{main5}
\end{eqnarray}
where $H\in\la_1$, $I\in\la_2$, $\Psi\in\la_{\rm ev}$, $A\in\la_1$, $F\in\la_2$,
$J\in\la_1$, $m$ and $\alpha$ are constants. These values have the
following physical interpretation: $\Psi$ is the tensor wave function of
electron, $A$ and $F$ are the potential and strength of electromagnetic
field respectively, $J$ is the electric current generated by the electron,
$m$ is the electron mass, and $\alpha$ is a real constant dependent on
physical units (the speed of light is equal to 1).

It can be shown (see \cite{Marchuk:Cimento2})
that if an even exterior form $\Psi$ satisfies (\ref{main2}),
then the 1-form $J=\Psi H\Psi^*$ satisfies the equality
\begin{equation}
\delta J=0,
\label{charge:conservation}
\end{equation}
which is called {\it a charge conservation law} for the Dirac type tensor
equation. Taking into account the identity $\delta^2=0$, we see that
(\ref{charge:conservation}) is consistent with equation (\ref{main4}).

\theorem 4. System of equations (\ref{main1}-\ref{main5}) is invariant
under the following gauge transformation
\begin{eqnarray*}
\Psi &\to& \Psi^\prime=\Psi\,\exp(\lambda I),\\
A &\to& A^\prime=A-d\lambda,
\end{eqnarray*}
where $\lambda=\lambda(x)$ is a smooth scalar function, i.e.,
$\lambda\in\la_0$ and $\exp(\lambda I)=\cos\,\lambda + I\sin\,\lambda$.
\par

\proof. Let us multiply equation (\ref{main2}) from the right by
$\exp(\lambda I)$. Then, using the relations $[H,I]=0$ and
$d-\delta=e^\mu\partial_\mu$, we  obtain
$$
(d-\delta)\Psi^\prime+A^\prime \Psi^\prime I+m\Psi^\prime HI=0.
$$
The identities
\begin{equation}
\exp(\lambda I)^*=\exp(-\lambda I)=\exp(\lambda I)^{-1}
\label{U1}
\end{equation}
give the invariance of the equality
$$
J=\Psi H\Psi^*=\Psi^\prime H(\Psi^\prime)^*.
$$
This completes the proof.
\medskip

Identities (\ref{U1}), together with the identity
$$
\exp(\lambda I)^*=\exp(\lambda I)^\dagger,
$$
show that the set of even exterior forms
$$
\{\exp(\lambda I)\,:\,\lambda\in\la_0\},
$$
considered at any point $x\in\R^{1,3}$, is isomorphic to the
unitary group ${\rm U}(1)$. Theorem 4 corresponds to the well known
fact that the Dirac equation is invariant under the gauge transformations
from the group ${\rm U}(1)$.

Finally, consider the Lagrangian (Lagrange density)
$$
\L=\frac{1}{4}\tr(H(\Psi^* Q+Q^*\Psi)),
$$
where
$$
Q= (d-\delta)\Psi I-A\Psi-m\Psi H.
$$
Variating the Lagrangian $\L$ w.r.t. components of the exterior form $\Psi$,
we may derive the Dirac type tensor equation.


\end{document}